\documentclass{optica-article}

\journal{opticajournal} 

\articletype{Research Article}

\usepackage{lineno}

\begin{document}

\title{Parallel compressive super-resolution imaging with wide field-of-view based on physics enhanced network}

\author{Xiao-Peng Jin,\authormark{1} An-Dong Xiong,\authormark{1} Wei Zhang,\authormark{1} Xiao-Qing Wang,\authormark{2} Fan Liu,\authormark{2} Chang-Heng Li,\authormark{2} Xu-Ri Yao,\authormark{1,3} Xue-Feng Liu,\authormark{2,4,6} and Qing Zhao\authormark{1,3,5}}

\address{\authormark{1}Center for Quantum Technology Research and Key laboratory of Advanced Optoelectronic Quantum Architecture and Measurements (MOE), School of Physics, Beijing Institute of Technology, Beijing 100081, China\\
	\authormark{2}Key Laboratory of Electronics and Information Technology for Space Systems, National Space Science Center, Chinese Academy of Sciences, Beijing 100090, China\\
	\authormark{3}Beijing Academy of Quantum Information Sciences, Beijing 100193, China\\
	\authormark{4}University of Chinese Academy of Sciences, Beijing 100049, China}

\email{\authormark{5}qzhaoyuping@bit.edu.cn}
\email{\authormark{6}liuxuefeng@nssc.ac.cn} 


\begin{abstract*}
	Achieving both high-performance and wide field-of-view (FOV) super-resolution imaging has been attracting increasing attention in recent years. However, such goal suffers from long reconstruction time and huge storage space. Parallel compressive imaging (PCI) provides an efficient solution, but the super-resolution quality and imaging speed are strongly dependent on precise optical transfer function (OTF), modulation masks and reconstruction algorithm. In this work, we propose a wide FOV parallel compressive super-resolution imaging approach based on physics enhanced network. By training the network with the prior OTF of an arbitrary $128 \times 128$-pixel region and fine-tuning the network with other OTFs within rest regions of FOV, we realize both mask optimization and super-resolution imaging with up to $1020 \times 1500$ wide FOV. Numerical simulations and practical experiments demonstrate the effectiveness and superiority of the proposed approach. We achieve high-quality reconstruction with $4 \times 4$ times super-resolution enhancement using only three designed masks to reach real-time imaging speed. The proposed approach promotes the technology of rapid imaging for super-resolution and wide FOV, ranging from infrared to Terahertz.
\end{abstract*}

\section{Introduction}
\noindent Parallel compressive imaging (PCI) is an emerging super-resolution imaging method which derives from single-pixel imaging (SPI). In PCI, the objects are first modulated by high-resolution masks loaded on spatial modulators such as digital micromirror device (DMD), then captured by a low-resolution detector such as charge coupled device (CCD) and recovered by super-resolution algorithms eventually. Compared with SPI, PCI offers rapid imaging speed and efficient data-processing, hence has advantages on wide FOV scenarios. Meanwhile, it provides a solution for high-resolution imaging in situations where the detector pixel number is relatively low, thus, has been broadly used in divergent imaging fields. However, currently high-quality PCI with wide FOV is still difficult, restricting its practical applications.

In PCI, the optical transfer function (OTF), modulation masks and reconstruction algorithms directly determine the reconstruction quality, which are indispensable for achieving high-quality wide FOV super-resolution imaging. OTF denotes the pixel-relationship between detector and DMD, it is the premise for the reconstruction. In traditional PCI, researchers usually split the system into multiple independent SPI systems and recover them separately. In such cases, the influence of OTF is ignored. However, in the PCI system with wide FOV, the DMD and imaging lens need to be placed parallel to obtain clear images on the modulation plane. As a result, the non-parallel between DMD and detector breaks the ideal relationships and further results in poor imaging quality and resolution. Therefore, before imaging the actual target, the OTF measurement process is inevitable. Benefit from our previous work, we can efficiently and accurately calculate the OTF for the entire modulation region of the system.

Modulation masks and reconstruction algorithms are critical factors for the imaging quality and efficiency. There are many notable researches in these aspects, such as various Hadamard and Fourier matrix sampling strategies for modulation mask design; ghost imaging algorithms, Fourier algorithms and compressed sensing algorithms for reconstruction. However, in the case of low-sampling number, the above methods will cause the information loss of original image, leading to the decrease in quality and resolution.

Recently, the emergence of deep learning has enabled the integration of modulation mask and reconstruction algorithm optimization in SPI. The masks can be involved into well-designed network layers, and the advanced deep learning algorithms can be used to achieve high-quality reconstruction at extreme low sampling rate. Quite recently, Wang et al. proposed a physics enhanced deep learning approach for visible SPI, which is proved to be generalizable by blending a physics-informed layer and fine-tune process. Such methods have been extensively applied in SPI. However, in PCI with wide FOV, there are two challenges associated with deep learning. First, when the real physical process is combined with network, i.e. the actual OTF is used, the low-resolution images captured by detector in PCI result in the significant huge data volume compared with SPI, which can easily expand the network and reach the limit of hardware. Second, even if we split the whole wide FOV into multiple small regions, the divergence between OTFs of these regions will lead to totally different networks, which need to be trained independently and therefore is extremely time-consuming. Due to the above difficulties, to the best of our knowledge, no deep learning-based PCI approach with consideration of the actual physical conditions has been identified.

In this work, we propose a wide FOV parallel compressive super-resolution imaging approach based on physics enhanced network. By first training both the modulation masks and network parameters of an arbitrary  $128 \times 128$-pixel DMD region with corresponding prior OTF which accomplish the physical enhancement, then fine-tuning the network for the rest DMD regions of the entire FOV, we not only achieve the high-quality and super-resolution reconstruction with wide FOV at low sampling number but also dramatically shorten the reconstruction time and network size. Numerical simulations and practical experiments are performed to demonstrate the effectiveness and superiority of the proposed method.

\section{Methods}

\begin{figure}[ht]
	\centering\includegraphics[width=1.0\linewidth]{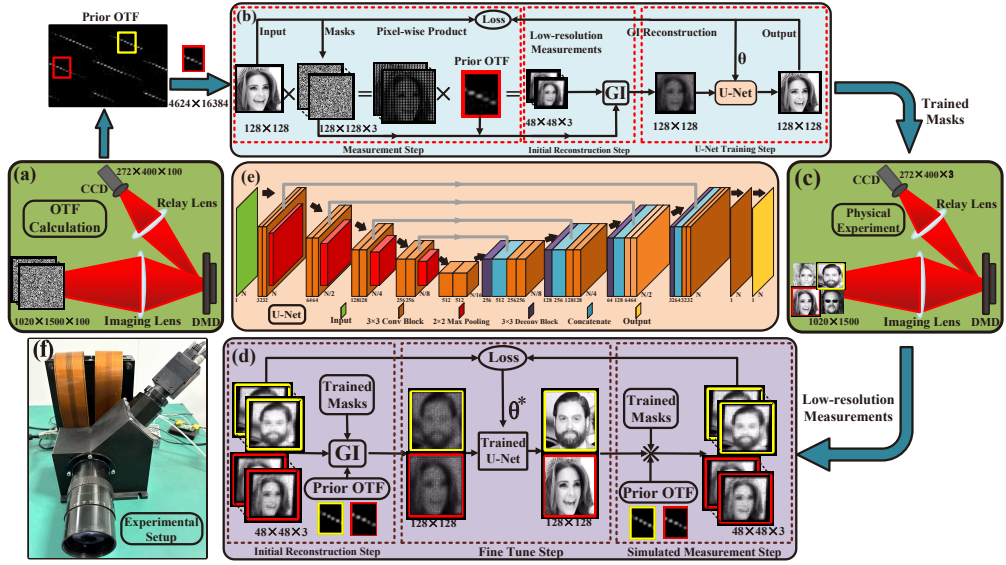}
	\caption{Schematic diagram of proposed approach. (a) OTF calculation process. (b) Network training part. (c) Physical experiment part. (d) Network fine-tuning part. (e) Diagram of the U-Net structure. (f) Actual experimental prototype.}
	\label{fig:1}
\end{figure}

As illustrated in Fig.1, the proposed approach consists of three parts: network training part, physical experiment part, and network fine-tuning part.

The network training part simulates the actual physical process of a random selected small DMD region to generate a set of trained modulation masks and network parameters. As the indispensable of OTF for PCI, thus, before network training, the OTF of whole DMD region should be calibrated to provide prior knowledge of the system, which is shown in Fig.1(a). 100 random binary masks are used for the calibration and the details of the process are elaborated in our previous work. The network training part contains three steps, as shown in Fig. 1(b). The first step is measurement step, which utilizes OTF of a random selected region to simulate the parallel measurement process. When we modulate the object $X\left( \varphi  \right)$ with size of $P \times Q$ in the selected region $\varphi $ by a set of $N$ masks $M$, the corresponding low-resolution images $Y\left( \varphi  \right)$ with size of $p \times q$ captured by detector can be expressed as:
\begin{equation}
\begin{aligned}
Y\left( \varphi  \right) = PCI\left( {C\left( \varphi  \right),M,X\left( \varphi  \right)} \right) = \left\{ {y{{\left( \varphi  \right)}_{m,i}}} \right\} 
= \left\{ {C{{\left( \varphi  \right)}_i} * col\left( {{M_m} \times X\left( \varphi  \right)} \right) + nois{e_{m,i}}} \right\},
\end{aligned}
\label{eq:refname1}
\end{equation}
where ${M_m}$ denotes the $m$th mask, $y{\left( \varphi  \right)_{m,i}}$ denotes the $i$th pixel value of the column-wise vectorized version for $m$th low-resolution image with $1 \le m \le N$, $nois{e_{m,i}}$ denotes the corresponding measurement noise, $C{\left( \varphi  \right)_i}$ is the $i$th row of the corresponding OTF with $\varphi $, which represents the contribution of each pixel on DMD to a given detector pixel $i$, $1 \le i \le \left( {p \times q} \right)$, $ \times $ denotes pixel-wise product, $ * $ denotes matrix product, $col\left(  \cdot  \right)$ denotes to form a column-wise vector. Distinct from Convolution Layer in SPI, the parallel measurement process integrates the OTF and is fulfilled by the Multiply Layer owing to it captures low-resolution images instead of single-pixel values. The second step is initial reconstruction step based on GI algorithm, which combines low-resolution images, masks and OTF to obtain an initial reconstruction of object ${X_{GI}}\left( \varphi  \right)$, 
\begin{equation}
\begin{aligned}
& {X_{GI}}\left( \varphi  \right) = GI\left( {C\left( \varphi  \right),M,Y\left( \varphi  \right)} \right)= Reshape \\ & \left\{ {\frac{1}{{\left( {p \times q} \right)}}\sum\limits_{i = 1}^{p \times q} {\sum\limits_{m = 1}^N {y{{\left( \varphi  \right)}_{m,i}}\left( {C\left( \varphi  \right)_i^T \times \left( {col\left( {{M_m}} \right)} \right)} \right)} } } \right\},
\end{aligned}
\label{eq:refname2}
\end{equation}
where $Reshape\left(  \cdot  \right)$ denotes the matrix deformation to the size of $P \times Q$. The third step is U-Net training step, as shown in Fig.~\ref{fig:1}(e), which includes four Down-sampling Layers, four Up-sampling Layers and four Concatenate Layers. It takes ${X_{GI}}\left( \varphi  \right)$ as input and produces an estimation ${X_{out}}\left( \varphi  \right) = {U_\theta }\left( {{X_{GI}}\left( \varphi  \right)} \right)$ as output. So far, the input training image $X\left( \varphi  \right)$ is forward propagated from ${X_{GI}}\left( \varphi  \right)$ to ${X_{out}}\left( \varphi  \right)$ through the whole network training part. Then, the modulation masks and U-Net parameters are optimized by solving the loss function between $X\left( \varphi  \right)$ and ${X_{out}}\left( \varphi  \right)$,
\begin{equation}
\begin{aligned}
\left\{ {{U_{{\theta ^ * }}},{M^ * }} \right\} &= \mathop {\arg \min }\limits_{\theta ,M} {\left\| {{X_{out}}\left( \varphi  \right) - X\left( \varphi  \right)} \right\|^2} \\ &= \mathop {\arg \min }\limits_{\theta ,M} {\left\| {{U_\theta }\left( {{X_{GI}}\left( \varphi  \right)} \right) - X\left( \varphi  \right)} \right\|^2}.
\end{aligned}
\label{eq:refname3}
\end{equation}
Compared with traditional PCI, the proposed approach incorporates the real OTF as prior knowledge, which is more consistent with the actual situation and is more conducive to high-quality reconstruction.

With the trained masks ${M^ * }$, we can measure the practical objects and capture the low-resolution measurements ${Y^ * }\left( \mu  \right) = PCI\left( {C\left( \mu  \right),{M^ * },X\left( \mu  \right)} \right)$
through physical experiment part, which is a PCI system with wide FOV shown in Fig.~\ref{fig:1}(c) for diagram and Fig.~\ref{fig:1}(f) for real prototype. $X\left( \mu  \right)$ is the object in another regions $\mu $ along the entire wide FOV with the same size of object $X\left( \varphi  \right)$, and $C\left( \mu  \right)$ is the corresponding OTF. The different values between $C\left( \mu  \right)$ and $C\left( \varphi  \right)$ will lead to the network mismatch, which will cause the inaccurate for reconstruction of $X\left( \mu  \right)$. However, benefit from the similar structure of OTFs in different regions in one optical system, we can increase the network generalization through the fine-tune process.

The network fine-tuning part is applied to adjust the U-Net parameters for different DMD regions to increase the network generalization and further improve the reconstruction quality of the full FOV range. It consists of three steps which are partially similar with counterparts in network training part, as shown in Fig.~\ref{fig:1}(d). The first step is initial reconstruction step based on GI algorithm, which generates an initial reconstruction result $X_{GI}^ * \left( \mu  \right) = GI\left( {C\left( \mu  \right),{M^ * },{Y^ * }\left( \mu  \right)} \right)$, similarly, according to Eq. (2). The second step is network fine-tuning step, which takes $X_{GI}^ * \left( \mu  \right)$ as input and produces a reconstructed image $X_{out}^ * \left( \mu  \right) = {U_{{\theta ^ * }}}\left( {X_{GI}^ * \left( \mu  \right)} \right)$ as output through the trained U-Net. The third step is simulated measurement step, which calculates the measurements of $X_{out}^ * \left( \mu  \right)$ under the same condition with trained masks and OTF, ${Y^{ *  * }}\left( \mu  \right) = PCI\left( {C\left( \mu  \right),{M^ * },X_{out}^ * \left( \mu  \right)} \right)$. Then, part of the network parameters is fine-tuned by solving the loss function between simulated measurements and practical measurements,
\begin{equation}
\begin{aligned}
\left\{ {{U_{{\theta ^{ *  * }}}}} \right\} = \mathop {\arg \min }\limits_{{\theta ^ * }} {\left\| {{Y^ * }\left( \mu  \right) - {Y^{ *  * }}\left( \mu  \right)} \right\|^2}.
\end{aligned}
\label{eq:refname4}
\end{equation}
After fine-tuning the network parameters, a high-quality reconstruction $X_{out}^ * \left( \mu  \right)$ can be obtained eventually.

Through the proposed approach, the reconstruction of the whole FOV is converted into the network training for a random selected small region $\varphi $ and the network fine-tuning for all the rest regions $\mu $. Assuming the whole FOV is divided into $n$ small regions, each small region needs ${T_1}$ time for network training and ${T_2}$ time for network fine-tuning, our approach only needs $({T_1} + n \cdot {T_2})$ in total against the $(n \cdot {T_1})$ for training each region independently. In this way, we not only greatly reduce the network storage but also promote the sampling-reconstruction efficiency. Furthermore, the prior knowledge of OTF introduces the real physical process into the network and can significantly improve the quality of reconstruction.

\section{Experiment and discussion}
\subsection{System description}
The actual PCI system used in this work is shown in Fig.1~(f). Light source is natural ambient light, and the illuminated target is projected onto the DMD through a zoom imaging lens. The resolution of the DMD is $1280 \times 1920$ (DLP9500, TI, USA) with each micro-mirror of 10.8$\mu m$. The target image on DMD is successively modulated by the trained masks. Afterward, the modulated high-resolution images are projected onto the detector (MV-CA017-100M) through the relay lens, with the theoretical under-sampling factor of $\left( {4 \times 4} \right):1$.  As mentioned above, the non-parallel between DMD and detector makes imaging with precise alignment between DMD and detector pixels impossible. Therefore, the imaging system of the relay lens is designed to satisfy the Scheimpflug principle for a large clear imaging area of the whole FOV. In actual experiments, the sizes of DMD and detector in use are $1020 \times 1500$ and $272 \times 400$ pixels, respectively, which is slightly larger than $255 \times 375$ pixels according to the theoretical under-sampling factor because of the distortion. Using our previously proposed work, we measure the OTF of the above FOV region, which size is $\left( {272 \times 400} \right):\left( {1020 \times 1500} \right)$.

\subsection{Numerical simulation}
We first validate the proposed approach via numerical simulation based on the actual OTF of the experimental system. For the network training part, we select a $128 \times 128$ region "A" located in the center of DMD, which is projected onto detector with $48 \times 48$-pixel, thus the OTF of this region is $\left( {48 \times 48} \right):\left( {128 \times 128} \right)$. The dataset used for training is ${128 \times 128}$-pixel images from CelebAMask-HQ, which is split into training set, validation set and test set with ratios of $27000:2970:30$. The modulation masks are three binary matrices with $128 \times 128$-pixel which are reduplicated from the $4 \times 4$-pixel elements considering the theoretical under-sampling factor of  $\left( {4 \times 4} \right):1$. Therefore, the sampling rate is approximately $3/16$. The noise in physical measurement step is set as Gaussian noise for increasing the network stability, $noise = {\sigma ^2} \times mean\left( y \right) \times N\left( {0,1} \right)$, where $mean\left( y \right)$ is the mean value of original measurements, $N\left( {0,1} \right)$ denotes the standard normal distribution, and $\sigma $ is the standard deviation with $\sigma  = 0.3$ in this work. For the implementation of the network, the learning rate is set to 0.0002, and the batch size is 15. For the physical experiment part, we also simulate the measurement process for objects in test set together with a digital resolution chart, and the Gaussian noise is also introduced to verify the feasibility and effectiveness with $\sigma  = 0,0.3,0.5$. For the network fine-tuning part, we only fine-tune the parameters in first three Convolution layers. The training was conducted in a computer with AMD Ryzen 7 5800 8-Core Processor @ 3.40GH, 16GB RAM, an NVIDIA RTX 3060 GPU, which converged within 30 epoches after 5 hours.

To quantitatively evaluate the reconstructed image quality, we adopt the peak signal-to-noise ratio (PSNR) and the structural similarity (SSIM), 
\begin{equation}
\begin{aligned}
PSNR = 10\log \left( {\frac{{{{\left( {{2^n} - 1} \right)}^2}}}{{\sum\nolimits_{i,j = 1}^{M,N} {{{\left[ {Y\left( {i,j} \right) - X\left( {i,j} \right)} \right]}^2}} }}} \right),
\end{aligned}
\label{eq:refname5}
\end{equation}

\begin{equation}
\begin{aligned}
SSIM = \frac{{\left( {2{\mu _X}{\mu _Y} + {C_1}} \right)\left( {2{\sigma _{XY}} + {C_2}} \right)}}{{\left( {\mu _X^2 + \mu _Y^2 + {C_1}} \right)\left( {\sigma _X^2 + \sigma _Y^2 + {C_2}} \right)}},
\end{aligned}
\label{eq:refname6}
\end{equation}

where $n$ is the bit number of the detector which is 16 in our experiment, $X$ and $Y$ are the original and reconstructed images, respectively, ${\mu _X}$ and ${\sigma _X}$ represent the average value and the variance of $X$, ${\sigma _{XY}}$ denotes the covariance of $X$ and $Y$, ${C_1}$ and ${C_2}$ are constants used to maintain stability.

Fig.3 shows the simulated imaging results with targets located on two different DMD regions. For groups of columns \uppercase\expandafter{\romannumeral1}-\uppercase\expandafter{\romannumeral6} to the target in region “A” and for columns \uppercase\expandafter{\romannumeral7}-\uppercase\expandafter{\romannumeral12} the target is on the top right corner of DMD named “B”. Columns \uppercase\expandafter{\romannumeral1} and \uppercase\expandafter{\romannumeral2} are the ground truth of test images and measured low-resolution images of the detector. \uppercase\expandafter{\romannumeral3}-\uppercase\expandafter{\romannumeral6} are reconstructions of GI, compressed sensing algorithm TVAL3, our approach without fine-tuning (W/O FT) part and with fine-tuning (W/FT) part, respectively. Columns \uppercase\expandafter{\romannumeral7}-\uppercase\expandafter{\romannumeral12} are the counterparts of target in region ‘B’, and the corresponding PSNR and SSIM values are listed in Table 1. 

For the test image of a human face in region “A”, we observe that the structural noise severely decreases the GI reconstruction quality, which is shown in column \uppercase\expandafter{\romannumeral3} TVAL3 and “W/O FT” can achieve super-resolution reconstruction for overall image size with better qualities, but the details such as letters of “TRY” in the left top corner are still obscure. Besides, due to the noise added during the training part, the U-net is robustness against noise, therefore the PSNR and SSIM of “W/O FT” degrade more slowly than TVAL3 when the noise increases, which can be seen in Table.1. The “W/FT” outperforms all other methods according to PSNR, SSIM and visual effect. Despite the quality is also inversely proportional with noise, the letters of “TRY” can be recovered with high-fidelity through fine-tuning the measurements via formula (4). The outline of letter “T” is distinguished even under the noise level of 0.5. For the resolution chart with low similarity of the training database, the regularities in the above results also exist. As shown in the corresponding partial enlargements of red boxes in Fig.3, a part of narrow stripes can be recovered by fine-tuning process against the other methods, which indicates the proposed approach is generalizable for image reconstruction.

For the image in region “B”, the most significant difference compared with imaging results of region “A” is the “W/O FT” method, which contains the unattractive noise. This is reasonable as the region is changed with that in the U-net training part, thus the dissimilarity between the corresponding OTFs seriously deteriorates the reconstruction quality. However, by introducing the changed OTF of region “B” into fine-tuning part, we can finally obtain a high-quality super-resolution reconstruction, which can be clearly seen in column \uppercase\expandafter{\romannumeral12} and enlarged details in green boxes of Fig.3. From the performance values in Table.1, we demonstrate the proposed approach can achieve high-quality super-resolution imaging for various regions in the wide FOV via only training an arbitrarily selected region and fine-tuning the rest areas, which also verifies the effectiveness and superiority of our proposed method. Moreover, the reconstruction for fine-tuning part just needs 1s against 15s for TVAL3, which has potential for real-time imaging in PCI.

\begin{figure}[ht!]
	\centering\includegraphics[width=1.0\linewidth]{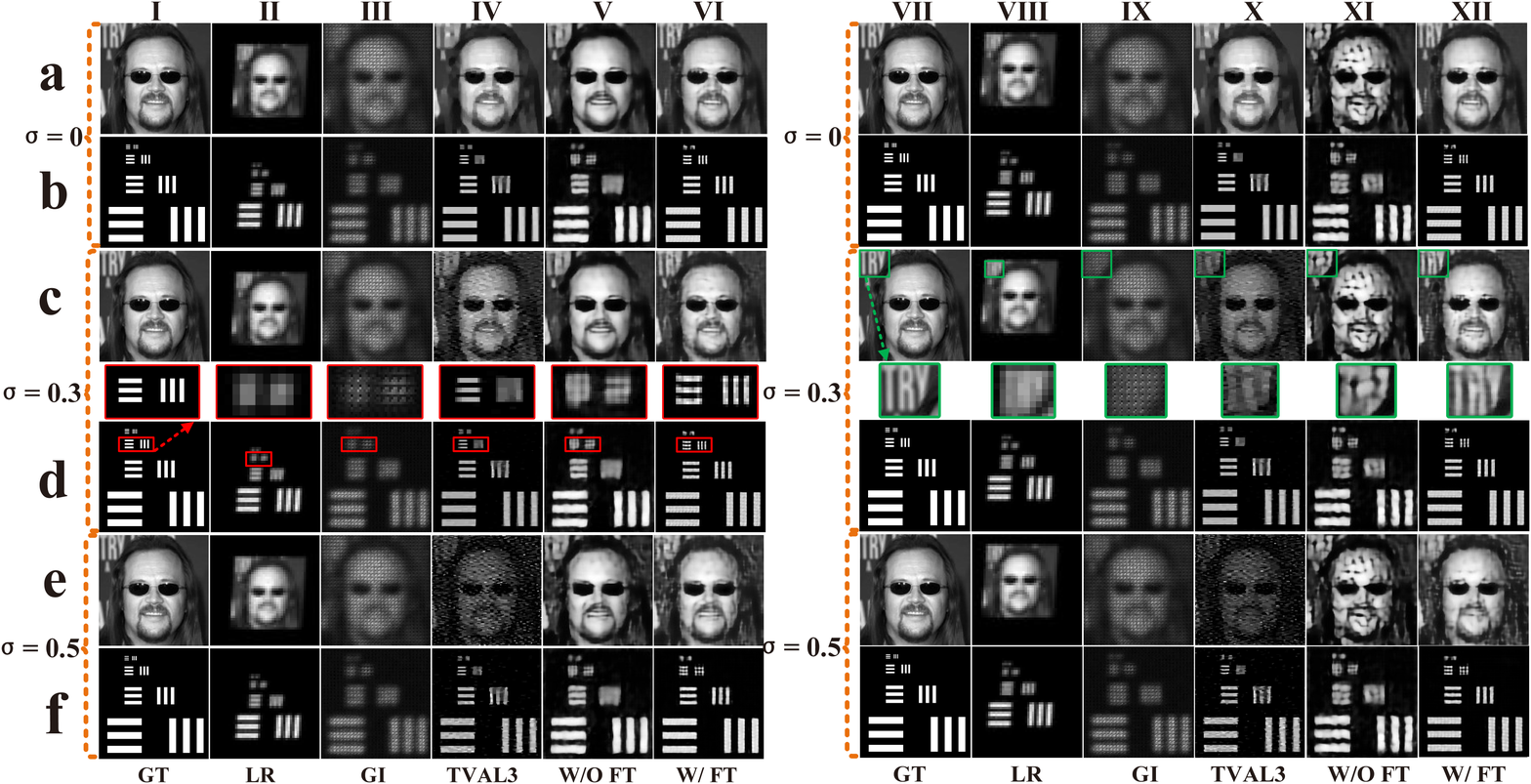}
	\caption{Simulation results and partial enlargements for targets located on two different DMD regions under three noise levels. (\uppercase\expandafter{\romannumeral1}-\uppercase\expandafter{\romannumeral2}) Ground truth and measured low-resolution images. (\uppercase\expandafter{\romannumeral3}-\uppercase\expandafter{\romannumeral6}) Reconstructions of GI, TVAL3, our approach without fine-tuning (W/O FT) part and with fine-tuning (W/FT) part, respectively. )(\uppercase\expandafter{\romannumeral7}-\uppercase\expandafter{\romannumeral12}) The counterparts of targets in region “B”.}
	\label{fig:2}
\end{figure}

\subsection{Practical experiment}
We then verify our approach for the digital image in real PCI system. The object is generated by repeating a $128 \times 128$ pixel binary image to $1020 \times 1500$ DMD region and padding zeros in margin, with the modulated masks are still the three trained masks as mentioned above. Figs.3(a-b) show the original digital object and low-resolution image obtained by direct observation of the detector without a coded mask. Fig.3(c-f) shows the reconstruction under different algorithms for a random region in the DMD. The corresponding PSNR and SSIM values are marked in orange in the images.

From the results, we observe the performances for each method conform to the aforementioned principles. Result of TVAL3 is affected by environment noise and has a high reconstruction error. The quality of “W/O FT” result is higher, but the changed OTF of imaging region destructs the accuracy of physical model and restricts achieving better reconstruction. The “W/ FT” approach further improves the reconstruction quality in terms of visual effect and evaluated index, with the letters “TRY” identified, and the noise eliminated. Above all, for wide FOV digital object, the proposed approach can realize high-quality super-resolution imaging.

\begin{figure}[ht]
	\centering\includegraphics[width=1.0\linewidth]{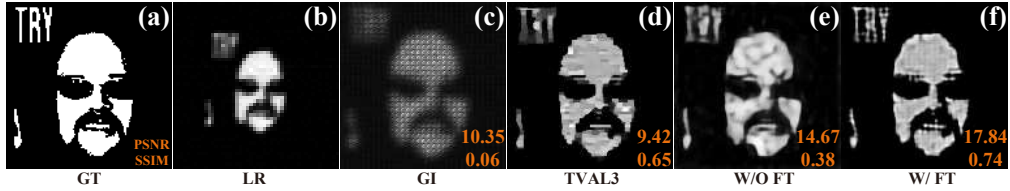}
	\caption{Experimental results for digital target. (a) Ground truth. (b) The low-resolution images on detector. (c-e) Reconstructions of GI, TVAL3, our approach without fine-tuning (W/O FT) part and with fine-tuning (W/FT) part, respectively.}
	\label{fig:3}
\end{figure}

We finally use our PCI system to image wide FOV real objects in a scenario. The trained masks are expanded to $1020 \times 1500$ DMD region. Fig. 4 shows the results of four different DMD regions “a-d”. Column \uppercase\expandafter{\romannumeral1} shows the low-resolution 48x48 pixel images directly observed by the detector without a coded mask. \uppercase\expandafter{\romannumeral2}-\uppercase\expandafter{\romannumeral5} are reconstructions of GI, TVAL3, “W/O FT” and “W/FT” with the size of $128 \times 128$, respectively. The mean values of pixels along each row in dashed boxes and each line in solid boxes for the measured and reconstructed results are shown under the corresponding images.

\begin{figure}[ht]
	\centering\includegraphics[width=1.0\linewidth]{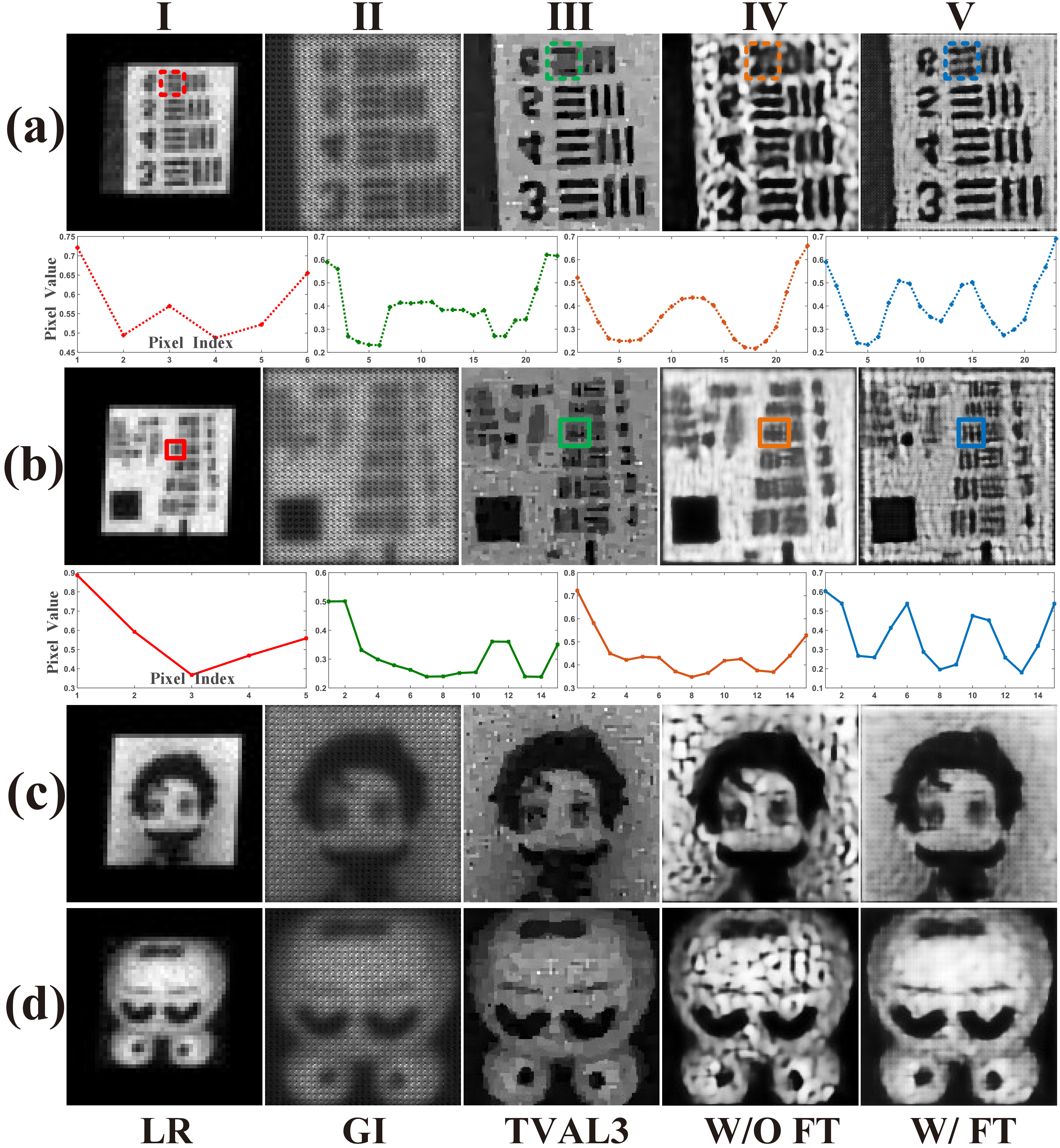}
	\caption{Experimental results for real objects and mean values of pixels along each row in dashed boxes and each line in solid boxes. (a-d) Four different regions in a scenario. (\uppercase\expandafter{\romannumeral1}-\uppercase\expandafter{\romannumeral4}) Low-resolution images captured on detector and reconstructions of TVAL3, our approach without fine-tuning (W/O FT) part and with fine-tuning (W/FT) part, respectively.}
	\label{fig:4}
\end{figure}

As shown in Fig.4, whether for the dolls or the resolution chart, the TVAL3 results suffer from severely noise brought by environment interference. The “W/O FT” results are simultaneously affected by changed OTFs, leading to the blurry visual effect and poor reconstruction qualities. Besides, the above two approaches can only improve the resolution in overall image size under low sampling number. The details like stripes in resolution chart cannot be distinguished by them. As shown in Fig. 4, the green and orange dashed lines lose some peaks, while the green and orange solid lines are too smooth. On the contrary, the “W/FT” method enables the super-resolution reconstruction and further improves the imaging visual effect, by which the indistinguishable parts in low-resolution images can be clearly separated. The red solid line in Fig. 4 is smooth but the blue solid line expresses obvious periodic variation, in which the peaks and troughs are resolvable. In addition, we prove the 4x4 times resolution enhancement by comparing the peak width of red dashed line and blue dashed line in Fig.4. Furthermore, the backgrounds for dolls of “W/FT” are uniform, which prevent the environmental noises. As the consequence, through our proposed approach, we can achieve high-quality super-resolution imaging for wide FOV objects, and the results are superior to the existing algorithms.

\section{Conclusion}
In this work, we propose a physics enhanced network approach for wide FOV parallel compressive super-resolution imaging. The approach consists of two mainly parts. First, by introducing the prior OTF of a randomly selected imaging region into network training part, we obtain the modulation masks and image reconstruction network. Second, by calculating the loss function of actual measurements and simulated measurements of different imaging regions, we fine-tune the network parameters and further improve the reconstruction quality of other regions to satisfy the requirement of wide FOV target imaging. Numerical simulations and practical experiments are performed to demonstrate the effectiveness and superiority of the proposed method. Three binary masks are trained in this work, and the results outperform the existing methods in visual effect and evaluation index.

It is generally believed that the parallel compressive imaging can significantly reduce the sampling number via splitting the imaging region into multiple uniform single-pixel imaging systems. Benefit from the deep-learning approach, the reconstruction quality can still stay in high-level even under extra-low sampling number. However, when facing the wide FOV targets, if we still use the existing approach, i.e., dividing the whole region into multiple single-pixel systems and treating them as a uniform network, the changed OTFs of various regions will cause the totally different network, which is time-consuming dramatically increases network training time for a new PCI system. To solve this problem, based on the similar structure of various OTFs, we apply the fine-tuning operation in our approach. In this way, we increase the network generalization and achieve the high-performance imaging in quality for wide FOV targets. Besides, the three trained masks improve the imaging speed and raise efficiency, thus the PCI system can be carried out in real time and maintain high standard for unstable environments during the short modulation time, which is applicable to many scenarios such as objects recognition, microscopy and astronomy.

In future work, we will focus on two main aspects. First, we will continue reducing the modulation masks by considering the temporal correlation of image sequences to further enhance the imaging speed. Second, we will choose new network model such as transformer to improve the robustness against noise, increase the imaging quality and resolution in details.


\bibliography{sample}

\begin{thebibliography}{10}
\newcommand{\enquote}[1]{``#1''}

\bibitem{1}
M.~F. Duarte, M.~A. Davenport, D.~Takhar, \emph{et~al.}, \enquote{Single-pixel
  imaging via compressive sampling,} {\protect\JournalTitle{IEEE Signal
  Processing Magazine}} \textbf{25}, 83--91 (2008).

\bibitem{2}
M.~P. Edgar, G.~M. Gibson, and M.~J. Padgett, \enquote{Principles and prospects
  for single-pixel imaging,} {\protect\JournalTitle{Nature Photonics}}
  \textbf{13}, 13--20 (2019).

\bibitem{3}
G.~M. Gibson, S.~D. Johnson, and M.~J. Padgett, \enquote{Single-pixel imaging
  12 years on: a review,} {\protect\JournalTitle{Opt. Express}} \textbf{28},
  28190--28208 (2020).

\bibitem{4}
L.~McMackin, M.~A. Herman, B.~Chatterjee, and M.~Weldon, \enquote{{A
  high-resolution SWIR camera via compressed sensing},} in \emph{Infrared
  Technology and Applications XXXVIII,}  vol. 8353 B.~F. Andresen, G.~F. Fulop,
  and P.~R. Norton, eds., International Society for Optics and Photonics (SPIE,
  2012), p. 835303.

\bibitem{5}
J.~Ke and E.~Y. Lam, \enquote{Object reconstruction in block-based compressive
  imaging,} {\protect\JournalTitle{Opt. Express}} \textbf{20}, 22102--22117
  (2012).

\bibitem{6}
A.~Mahalanobis, R.~Shilling, R.~Murphy, and R.~Muise, \enquote{Recent results
  of medium wave infrared compressive sensing,} {\protect\JournalTitle{Appl.
  Opt.}} \textbf{53}, 8060--8070 (2014).

\bibitem{7}
H.~Chen, M.~S. Asif, A.~C. Sankaranarayanan, and A.~Veeraraghavan,
  \enquote{Fpa-cs: Focal plane array-based compressive imaging in short-wave
  infrared,} in \emph{2015 IEEE Conference on Computer Vision and Pattern
  Recognition (CVPR),}  (2015), pp. 2358--2366.

\bibitem{8}
Z.~Wu and X.~Wang, \enquote{Focal plane array-based compressive imaging in
  medium wave infrared: modeling, implementation, and challenges,}
  {\protect\JournalTitle{Appl. Opt.}} \textbf{58}, 8433--8441 (2019).

\bibitem{9}
L.~Zhang, J.~Ke, S.~Chi, \emph{et~al.}, \enquote{High-resolution fast mid-wave
  infrared compressive imaging,} {\protect\JournalTitle{Opt. Lett.}}
  \textbf{46}, 2469--2472 (2021).

\bibitem{10}
J.~P. Dumas, M.~A. Lodhi, W.~U. Bajwa, and M.~C. Pierce, \enquote{Computational
  imaging with a highly parallel image-plane-coded architecture: challenges and
  solutions,} {\protect\JournalTitle{Opt. Express}} \textbf{24}, 6145--6155
  (2016).

\bibitem{11}
X.~Yuan and Y.~Pu, \enquote{Parallel lensless compressive imaging via deep
  convolutional neural networks,} {\protect\JournalTitle{Opt. Express}}
  \textbf{26}, 1962--1977 (2018).

\bibitem{12}
X.-P. Jin, A.-D. Xiong, X.-Q. Wang, \emph{et~al.}, \enquote{Long-distance
  mid-wave infrared super-resolution compressive imaging,}
  {\protect\JournalTitle{Optics \& Laser Technology}} \textbf{157}, 108740
  (2023).

\bibitem{13}
Y.~Cai, S.~Li, W.~Zhang, \emph{et~al.}, \enquote{A detail-enhanced sampling
  strategy in hadamard single-pixel imaging,} {\protect\JournalTitle{Chin. Opt.
  Lett.}} \textbf{21}, 071101 (2023).

\bibitem{14}
M.-J. Sun, M.~Tong, M.~Edgar, \emph{et~al.}, \enquote{A russian dolls ordering
  of the hadamard basis for compressive single-pixel imaging,}
  {\protect\JournalTitle{Scientific Reports}} \textbf{7}, 3464 (2017).

\bibitem{15}
W.-K. Yu, \enquote{Super sub-nyquist single-pixel imaging by means of
  cake-cutting hadamard basis sort,} {\protect\JournalTitle{Sensors}}
  \textbf{19} (2019).

\bibitem{16}
X.~Yu, R.~Stantchev, F.~Yang, and E.~Pickwell-MacPherson, \enquote{Super
  sub-nyquist single-pixel imaging by total variation ascending ordering of the
  hadamard basis,} {\protect\JournalTitle{Scientific Reports}} \textbf{10}
  (2020).

\bibitem{17}
Z.~Zhang, X.~Wang, G.~Zheng, and J.~Zhong, \enquote{Hadamard single-pixel
  imaging versus fourier single-pixel imaging,} {\protect\JournalTitle{Opt.
  Express}} \textbf{25}, 19619--19639 (2017).

\bibitem{18}
B.~I. Erkmen and J.~H. Shapiro, \enquote{Ghost imaging: from quantum to
  classical to computational,} {\protect\JournalTitle{Adv. Opt. Photon.}}
  \textbf{2}, 405--450 (2010).

\bibitem{19}
Y.~LeCun, Y.~Bengio, and G.~Hinton, \enquote{Deep learning,}
  {\protect\JournalTitle{Nature}} \textbf{521}, 436--44 (2015).

\bibitem{20}
G.~Barbastathis, A.~Ozcan, and G.~Situ, \enquote{On the use of deep learning
  for computational imaging,} {\protect\JournalTitle{Optica}} \textbf{6},
  921--943 (2019).

\bibitem{21}
C.~Higham, R.~Murray-Smith, M.~Padgett, and M.~Edgar, \enquote{Deep learning
  for real-time single-pixel video,} {\protect\JournalTitle{Scientific
  Reports}} \textbf{8} (2018).

\bibitem{22}
F.~Wang, C.~Wang, C.~Deng, \emph{et~al.}, \enquote{Single-pixel imaging using
  physics enhanced deep learning,} {\protect\JournalTitle{Photon. Res.}}
  \textbf{10}, 104--110 (2022).

\bibitem{23}
O.~Ronneberger, P.~Fischer, and T.~Brox, \enquote{U-net: Convolutional networks
  for biomedical image segmentation,} in \emph{Medical Image Computing and
  Computer-Assisted Intervention -- MICCAI 2015,}  N.~Navab, J.~Hornegger,
  W.~M. Wells, and A.~F. Frangi, eds. (Springer International Publishing, Cham,
  2015), pp. 234--241.

\bibitem{24}
S.~Nayar, V.~Branzoi, and T.~Boult, \enquote{Programmable imaging: Towards a
  flexible camera,} {\protect\JournalTitle{International Journal of Computer
  Vision}} \textbf{70}, 7--22 (2006).

\bibitem{25}
C.-H. Lee, Z.~Liu, L.~Wu, and P.~Luo, \enquote{Maskgan: Towards diverse and
  interactive facial image manipulation,} in \emph{2020 IEEE/CVF Conference on
  Computer Vision and Pattern Recognition (CVPR),}  (2020), pp. 5548--5557.

\bibitem{26}
Z.~Wang, A.~Bovik, H.~Sheikh, and E.~Simoncelli, \enquote{Image quality
  assessment: from error visibility to structural similarity,}
  {\protect\JournalTitle{IEEE Transactions on Image Processing}} \textbf{13},
  600--612 (2004).

\bibitem{27}
C.~Li, \emph{An efficient algorithm for total variation regularization with
  applications to the single pixel camera and compressive sensing} (Rice
  University, 2010).

\end{thebibliography}






\end{document}